\documentclass[amsmath,amssymb,aps,pra,groupedaddress,twocolumn]{revtex4-1}
\usepackage{graphicx}
\usepackage{subfigure}
\usepackage{dsfont}
\usepackage{bm}
\usepackage{mathrsfs}
\usepackage{hyperref}
\hypersetup{colorlinks=true, linkcolor=black, citecolor=black, urlcolor=black}
\usepackage{comment}
\usepackage{braket}
\usepackage{todonotes}

\begin{document}

\title{Generating and detecting topological phases with higher Chern number}

\author{Abhijeet Alase}
\email[Corresponding author: ]{abhijeet.alase1@ucalgary.ca}
\affiliation{Institute for Quantum Science and Technology, and Department of 
Physics and Astronomy, University of Calgary, Alberta T2N 1N4, Canada}

\author{David L. Feder}
\affiliation{Institute for Quantum Science and Technology, and Department of 
Physics and Astronomy, University of Calgary, Calgary, Alberta, Canada T2N 1N4}

\begin{abstract}
Topological phases with broken time-reversal symmetry and Chern number
$|\mathcal{C}|\ge 2$ are of fundamental interest, but it remains unclear
how to engineer the desired topological Hamiltonian within the paradigm of 
spin-orbit-coupled particles hopping only between nearest neighbours of a 
static lattice. We show that phases with higher Chern number arise when the 
spin-orbit coupling satisfies a combination of spin and spatial rotation 
symmetries. We leverage this result both to construct minimal two-band tight-binding 
Hamiltonians that exhibit $|\mathcal{C}|=2,3$ phases, and to show that 
the Chern number of one of the energy bands can be inferred from the particle
spin polarization at the high-symmetry crystal momenta in the Brillouin zone. 
Using these insights, we provide a detailed experimental scheme for the 
specific realization of a time-reversal-breaking topological phase with 
$|\mathcal{C}|=2$ for ultracold atomic gases on a triangular lattice subject to 
spin-orbit coupling. The Chern number can be directly measured using Zeeman 
spectroscopy; for fermions the spin amplitudes can be measured directly via 
time of flight, while for bosons this is preceded by a short Bloch oscillation. 
Our results provide a pathway to the realization and detection of novel 
topological phases with higher Chern numbers in ultracold atomic gases.
\end{abstract}

\maketitle

\section{Introduction}

Since the breakthrough provided by the quantum Hall effect, the systems that
display topological features in the absence of an external magnetic field
have been at the center stage in research in condensed matter physics. The
landmark discovery of topological insulators as well as quantum anomalous Hall
systems established spin-orbit (SO) coupling as the dominant mechanism behind
topological band inversion~\cite{qi11}. In contrast to time-reversal invariant
topological systems including two-dimensional (2D) topological insulators, protected phases
characterized by higher Chern numbers $|\mathcal{C}|\ge 2$ are theoretically
possible in time-reversal broken systems~\cite{ryu10}. (Note that in this work
the Chern number is understood to mean the first Chern number.) Such systems 
are predicted to show rich physics~\cite{,wangyao12,barkeshli12,cook14b,song15}
and in the solid state have potential applications such as in low-power 
consumption electronics~\cite{zhang12}.

Experimentally, only a handful of solid-state materials have been synthesized 
that exhibit $|\mathcal{C}|\ge 2$ phases~\cite{li20,chen20,ge20,Zhao2020}, though
photonic platforms have also met with some success~\cite{skirlo15}. One 
route to the realization of higher Chern number phases in the solid state is by 
the use of layered systems~\cite{trescher12,huang15,Zhao2020}, such as thin 
films of topological insulators~\cite{jiang12,wanglian13,cook14}, in which the 
Chern numbers of individual energy bands add up to yield an overall higher 
Chern number. Much of the theoretical 
research on higher Chern number bands has focused on model tight-binding 
Hamiltonians that host such phases~\cite{wangran11,sticlet2012,liu12,yang12,
lee15,peter15,huang15,slager2013,sticlet2013,chinghualee}. 
Most of these Hamiltonians either require high angular 
momentum bonding, next-nearest-neighbor (nnn) terms, or complicated geometries that 
are difficult to realize experimentally. For instance, the smallest known 
nearest-neighbor (nn) Hamiltonian that displays a $\mathcal{C}=2$ phase requires 
$d$ orbitals~\cite{cook16} or six bands~\cite{wangran11}. Identification of
solid-state material candidates with $|\mathcal{C}|\ge 2$ has generally relied
on {\it ab initio} calculations~\cite{cook14,cook16,li20,zhang20}. Despite their
importance, the mechanisms that lead to higher Chern number bands in realistic
systems remain poorly understood.

Ultracold atoms trapped in optical lattices hold the promise to realize novel 
and fascinating topological phases \cite{umucallar2008, Shao2008, li2008, 
zhang2010, zhao2011, alba2011, liu2013, abanin2013, zhu2013, dauphin2013, 
wang2013, hauke2014,  grusdt2014, wang2014,  deng2014, price2016, zhang2017, 
wang2018, atala2013, jotzu2014, duca2015,  aidelsburger2015, li2016,
flaschner2016, nakajima2016, wu2016, sun2018}. In contrast to solid-state 
systems, a diverse set of experimental techniques are available for controlling
and tuning the Hamiltonians governing ultracold atomic gases. For example, it
is straightforward to generate a wide variety of regular and non-Bravais 
lattices~\cite{Jo2012,Thomas2016,Gross2017}, including those with high or 
multiple orbital bands~\cite{Soltan2012,Li2016b}.
Recent experiments~\cite{wu2016,sun2018,wang2018} have shown that it is 
possible to realize SO coupling for ultracold atoms trapped in an 
optical lattice by using a Raman coherence technique~\cite{Lin2011}, but 
marrying the periodicity of the SO interactions to that of the 
underyling optical lattice is not straightforward for all geometries. To date,
only ${\mathcal C}=\pm 1$ topologically non-trivial phases have been realized 
experimentally in these systems.

The estimation of the Chern number in ultracold atom environments also remains 
a challenge, as routine transverse conductance measurements on solid-state 
systems are not straightforward to implement in ultracold atomic 
gases~\cite{liu2013,atala2013,grusdt2014,duca2015,zhang2017}. Whereas several 
approaches have been proposed~\cite{umucallar2008,Shao2008,li2008,zhang2010,
zhao2011,alba2011,liu2013,abanin2013,zhu2013,dauphin2013,wang2013,hauke2014,
grusdt2014,wang2014,deng2014,price2016,zhang2017} and 
implemented~\cite{atala2013,jotzu2014,duca2015,aidelsburger2015,li2016,
flaschner2016,nakajima2016,wu2016,sun2018} for the detection of topological 
order in ultracold atoms, most of them either do not apply to SO 
coupled systems or rely on experimentally challenging procedures such as 
mapping the Berry curvature over the entire Brillouin zone (BZ). 
Furthermore, distinguishing a topological phase with Chern number 
${\mathcal C}=\pm 2$ from a trivial one with ${\mathcal C}=0$ can be challenging
using approaches that are easier to implement 
experimentally~\cite{liu2013,zhang2017}. The difficulty
stems in part from the fact that weakly interacting bosons condense to a single 
point in the BZ at zero temperature, rather than occupying a full 
energy band like their fermionic counterparts, necessitating a dynamical 
approach to the detection of the topological order. Moreover, it 
appears that topological phases with Chern number ${\mathcal C}=\pm 2$ cannot 
be realized using ultracold atoms on a square lattice; for example, a proposed 
time-reversal-breaking two-band model~\cite{sticlet2012} requires strong 
next-nearest neighbor hopping.

In this work, we present two main results. First, we identify a simple 
mechanism that can give rise to higher values of the Chern number 
$|{\mathcal C}|\geq 2$ in individual bands, for time-reversal breaking systems. 
Second, we leverage the first result to propose an experimental scheme using 
ultracold atomic gases to realize a nearest-neighbor Hamiltonian that exhibits 
a topological phase with Chern number $|{\mathcal C}|=2$.  

The first result is the derivation of a relation between the Chern number
(mod~$2n$, $n\in\mathbb{Z}$) and the spin polarization at the high-symmetry
points in the BZ of a SO coupled Hamiltonian supported on a
Bravais lattice with $2n$-fold rotation symmetry $C_{2n}$. The key insight is a 
relation between the Chern number and the Berry-Zak phase of a closed loop that 
connects high-symmetry points of the BZ, Eq.~(\ref{chern-spin}). The relation 
can be extended to systems with any number of internal states, and to other 
non-Bravais lattices (i.e.\ Bravais lattices with attached basis) in a 
straightforward way.

The second result leverages the Chern-spin polarization relation to construct, 
among others, a minimal two-band nn tight-binding Hamiltonian on a triangular 
lattice that exhibits topological phases with $\mathcal{C} = \pm 2$,
Eq.~(\ref{tight-binding}). 
We provide a detailed proposal for realizing this 
model with ultracold atoms in optical lattices, and 
for the measurement of the lower-band Chern number. Our scheme judiciously 
combines the existing techniques for the generation of a triangular optical 
lattice~\cite{Becker_2010} with an extension of the SO coupling scheme 
implemented on a square lattice using Raman coherence~\cite{wu2016}. We design 
a novel scheme for the detection of the Chern number in bosonic systems using 
Bloch oscillations and Zeeman spectroscopy. Both these techniques are 
experimentally well established and widely employed~\cite{dahan96, morsch2001,
cristiani2002}. For fermionic systems, the Chern number can be determined
directly by time-of-flight (TOF) imaging~\cite{mckay2009}, without the need of 
Bloch oscillations.

Whereas the Hamiltonian that we construct and propose to realize 
satisfies a particular spin-space rotation symmetry, the resulting 
topological phase is protected against 
all particle number preserving perturbations regardless of whether they 
obey the aforementioned symmetry. This is a straightforward
consequence of the fact that the topological phase is characterized by 
the Chern number.
In this sense, the Hamiltonian 
we construct belongs to the class of Chern insulators,
and not crystalline topological insulators.

The paper is organized as follows. Section~\ref{sec:principle} discusses
a symmetry principle that can yield topological two-band Hamiltonians 
with higher Chern numbers for a given band. A key relation that links the spin
polarization at high-symmetry points in the BZ is derived in
Sec.~\ref{subsec:spin-chern}, and several examples of tight-binding Hamiltonians
with $|{\mathcal C}|\geq 2$ are given in Sec.~\ref{subsec:engineering}.
Section~\ref{sec:ultracold} is devoted to a detailed proposal for the 
implementation (Sec.~\ref{subsec:experiment}), its analysis 
(Secs.~\ref{sec:continuumHam} and \ref{subsec:tight}) and a detection scheme (Sec.~\ref{subsec:chern}) for 
a topological Hamiltonian on a triangular lattice with band Chern number 
$|{\mathcal C}|=2$. The results are summarized in 
Sec.~\ref{sec:conclusions}.

\section{Symmetry principle yielding higher Chern number}
\label{sec:principle}

\subsection{Chern-spin polarization relation}
\label{subsec:spin-chern}

In this section, we derive an important relationship between the Chern number
and the spin polarization at high-symmetry points in the BZ. Assume 
that the system is described by a gapped $2\times 2$ tight-binding Bloch 
Hamiltonian $H(\mathbf{k})$, with symmetry property
\begin{equation}
\label{symmetry}
    H(S\mathbf{k}) = U H(\mathbf{k}) U^\dagger, \quad U := e^{-i\pi m\sigma_z /2n}, 
\end{equation}
where $\sigma_z$ is the Pauli-$z$ matrix, and $m \in \{1,2,3\}$. 
The symmetry group is generated by counter-clockwise rotations around the 
$z$ axis by angle $\pi/n$; we denote the rotation in real space by 
$R := R_z(\pi/n)$ and in reciprocal space 
by $S := R_z(\pi/n)$. The symmetry 
condition for $m=1$ is satisfied by Rashba SO coupling, which for small 
$|\mathbf{k}|$ has the form $\mathbf{k} \times \vec{\sigma} \cdot \hat{z}$. The 
symmetry for $m=2$ ($m=3$) requires that the spin wave function of any one of 
the bands on the Bloch sphere rotates twice (thrice) as rapidly as the momentum 
vectors in the $xy$ plane of the BZ.

Let $\ket{u(\mathbf{k})}$ denote the lower band wavefunction of the Hamiltonian
$H(\mathbf{k})$. Let $\vec{\mathcal{A}}(\mathbf{k}) = 
\braket{u(\mathbf{k})|\vec{\nabla}_{\mathbf{k}} | u(\mathbf{k})}$
and $\mathcal{B}(\mathbf{k}) = \vec{\nabla}_{\mathbf{k}} \times \vec{\mathcal{A}}\cdot\hat{z}$
denote the Berry connection and the Berry curvature, respectively of the 
lower energy band. The Chern number 
\begin{equation}
\mathcal{C} = \frac{1}{2\pi i} \int_{\text{BZ}} \mathcal{B}(\mathbf{k})d^2\mathbf{k} 
\end{equation}
can be related to the 
spin wavefunctions at the high-symmetry momenta $\Gamma$ and $M$ in two steps 
using an approach similar to that discussed in Ref.\,\cite{fang12}. 
For any integer $\ell$, Eq.\,\eqref{symmetry} dictates
that the wavefunctions at crystal momenta $S^\ell\mathbf{k}$ 
and $\mathbf{k}$ satisfy 
$\ket{u(S^\ell\mathbf{k})} = e^{i\theta_\ell(\mathbf{k})}U^\ell \ket{u(\mathbf{k})}$ for
some real function $\theta_\ell$ on the BZ. 
Then 
\begin{align}
\mathcal{A}_i(S^\ell\mathbf{k}_0) & = 
\braket{u(\mathbf{q})|\partial_{q_i} | u(\mathbf{q})}\big\vert_{\mathbf{q} = S^\ell\mathbf{k}_0}\nonumber\\
    & = \braket{u(S^\ell\mathbf{k})|\sum_{j}S^{-\ell}_{ji}\partial_{k_j} 
    | u(S^\ell\mathbf{k})}\big\vert_{\mathbf{k} = \mathbf{k}_0} \nonumber\\
    & = \braket{u(\mathbf{k})|(U^\dagger)^{\ell}e^{-i\theta_\ell(\mathbf{k})}\sum_{j}
    S^{-\ell}_{ji}\partial_{k_j} 
    e^{i\theta_\ell(\mathbf{k})}U^{\ell}| u(\mathbf{k})}\nonumber\\
    & = \sum_{j}S^\ell_{ij}\braket{u(\mathbf{k})|(U^\dagger)^{\ell}e^{-i\theta_\ell(\mathbf{k})}\partial_{k_j} 
    e^{i\theta_\ell(\mathbf{k})}U^{\ell}| u(\mathbf{k})} \nonumber\\
    & = \sum_{j}S^\ell_{ij}\left(\mathcal{A}_j(\mathbf{k}_0) + i\partial_{k_j}\theta_\ell(\mathbf{k}_0)\right),
\end{align}
where $\partial_{k_i}\equiv\partial/\partial k_i$ is
defined for brevity. Therefore,
\begin{equation}
\label{berryconnection}
    \vec{\mathcal{A}}(S^\ell\mathbf{k}) = S^\ell\left(\vec{\mathcal{A}}(\mathbf{k}) +i\vec{\nabla}_{\mathbf{k}}\theta_\ell(\mathbf{k})\right).
\end{equation}
Because the unitary 
operator $U$ is independent of $\mathbf{k}$, the gauge independence of
the Berry curvature $\mathcal{B}(\mathbf{k})$ 
leads to 
\begin{align}
    \mathcal{B}(S^\ell\mathbf{k}) &= \vec{\nabla}_\mathbf{k}\times \vec{\mathcal{A}}(S^\ell\mathbf{k})\cdot\hat{z} \nonumber\\
    &= \vec{\nabla}_\mathbf{k}\times S^\ell
    \left(\vec{\mathcal{A}}(\mathbf{k}) +i\vec{\nabla}_{\mathbf{k}}\theta_\ell(\mathbf{k})\right)\cdot\hat{z}\nonumber\\
    &= \vec{\nabla}_\mathbf{k}\times \left(\vec{\mathcal{A}}(\mathbf{k}) +i\vec{\nabla}_{\mathbf{k}}\theta_\ell(\mathbf{k})\right)\cdot\hat{z}\nonumber\\
    &= \vec{\nabla}_\mathbf{k}\times \vec{\mathcal{A}}(\mathbf{k})\cdot\hat{z} = \mathcal{B}(\mathbf{k}),
\end{align}
where we have made use of the facts that $S^\ell$ is an orthogonal 
transformation, and that the curl of a gradient is zero. The result is 
$\mathcal{B}(S\mathbf{k}) = \mathcal{B}(\mathbf{k})$, where $\hat{z}$ denotes 
the unit vector in the $z$ direction. The loop encloses an area $A(\gamma)$, 
which covers exactly $1/2n$ of the total BZ and satisfies 
$\sum_{j=1}^{2n} S^jA(\gamma) = \operatorname{BZ}$. By invoking Stokes' 
theorem, one obtains
\begin{equation}
\label{zak-chern}
    \frac{\mathcal{C}}{2n} = \frac{1}{2\pi i}\int_{A(\gamma)} \mathcal{B}(\mathbf{k})d^2\mathbf{k} 
    = \frac{1}{2\pi i}\int_{\gamma} \vec{\mathcal{A}}(\mathbf{k})\cdot 
{d\mathbf{k}}\;(\mbox{mod}\,1),
\end{equation}
where\!$\pmod{1}$ accounts for the definition of the Berry-Zak phase acquired 
over a closed loop modulo $2\pi$.

\begin{figure}[t]
\centering
\includegraphics [width=0.9\columnwidth]{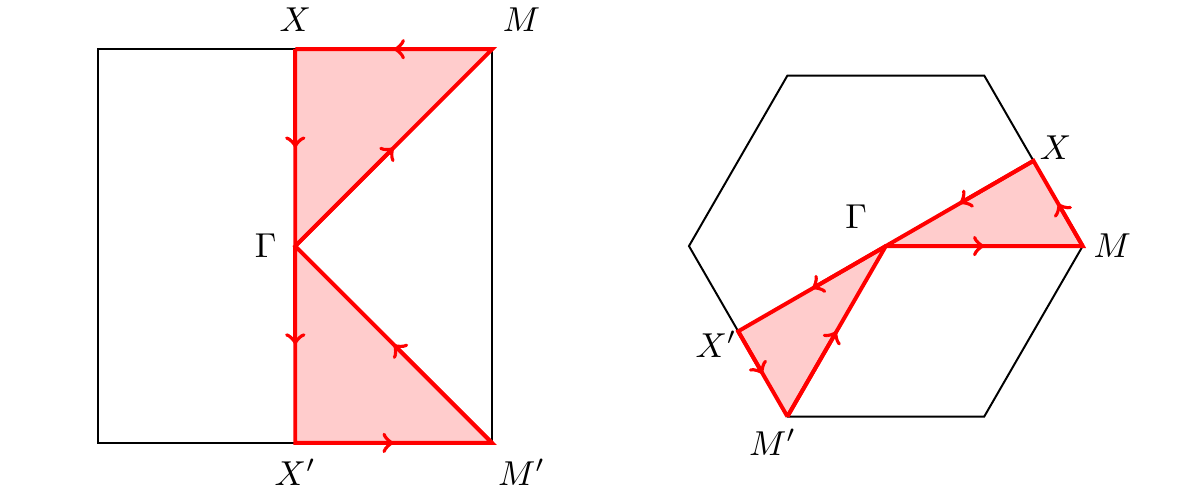}
\caption{(Color online) Loop $\gamma$ in the BZ of a two-dimensional Bravais 
lattice with $C_4$ (left) and $C_6$ (right) symmetry. The area $A(\gamma)$ 
enclosed by $\gamma$ is shaded in each case.
\label{fig:loops}}
\end{figure}

In the next step, the Berry phase acquired over the loop $\gamma$ is related to
the spin wavefunction at the points $\Gamma, M$, and $X$, as shown in 
Fig.~\ref{fig:loops}. The loop $\gamma$ may be decomposed into four parts,
\begin{multline}
\label{gammadecompose}
    {\gamma} 
     = (X \to \Gamma \to X') \cup  (X' \to M') \\
     \cup (M' \to \Gamma \to M) \cup (M \to X).
\end{multline}
The integral of the Berry connection over the second and the fourth segments on 
the right-hand side of Eq.\,\eqref{gammadecompose} cancel each other,
as the two segments correspond to the same paths in the BZ 
traversed in opposing directions. It is therefore only necessary to calculate 
the phases acquired over $X \to  \Gamma \to X'$ and $M' \to \Gamma \to M$.

Recall that on the
segment $M' \to \Gamma \to M$, the wave functions satisfy
$\ket{u(S^{n-1}\mathbf{k})} = e^{i\theta(\mathbf{k})}U^{n-1}\ket{u(\mathbf{k})}$,
where $\theta(\mathbf{k}) := \theta_{n-1}(\mathbf{k})$ is used for brevity. 
Because $M$ and $M'$ represent the same points in the BZ,
$\ket{u(M)} \propto \ket{u(M')}$, so that
\begin{equation}
\label{gauge}
    \theta(\mathbf{K}) = -\operatorname{Arg} \braket{u(\mathbf{K}) | U^{n-1} | u(\mathbf{K})},\quad
    \mathbf{K} \in \{\Gamma, M'\}.
\end{equation}
The Berry phase along the path $M' \to \Gamma \to M$ may be expressed as
\begin{equation}
\label{decompose}
    \int_{M' \to \Gamma \to M} \vec{\mathcal{A}}(\mathbf{k})\cdot {d\mathbf{k}} =
    \int_{\Gamma \to M} \vec{\mathcal{A}}(\mathbf{k})\cdot {d\mathbf{k}} -
    \int_{\Gamma \to M'} \vec{\mathcal{A}}(\mathbf{k})\cdot {d\mathbf{k}}.
\end{equation}
Using the symmetry properties, the first term on the right hand side of 
Eq.~\eqref{decompose} can be expressed as
\begin{align}
\label{connectionsymmetry}
    &\int_{\Gamma \to M} \vec{\mathcal{A}}(\mathbf{q})\cdot {d\mathbf{q}}
    = \int_{\Gamma \to M'} \vec{\mathcal{A}}(S^{n-1}\mathbf{k})\cdot S^{n-1}{d\mathbf{k}}\nonumber\\
    &\qquad = \int_{\Gamma \to M'} S^{n-1}(\vec{\mathcal{A}}(\mathbf{k}) +i\vec{\nabla}_{\mathbf{k}}\theta(\mathbf{k}))
    \cdot S^{n-1}{d\mathbf{k}}\nonumber\\
    &\qquad = \int_{\Gamma \to M'} (\vec{\mathcal{A}}(\mathbf{k}) +i\vec{\nabla}_{\mathbf{k}}\theta(\mathbf{k}))
    \cdot {d\mathbf{k}}\nonumber\\
    &\qquad = \int_{\Gamma \to M'} \vec{\mathcal{A}}(\mathbf{k})\cdot {d\mathbf{k}} + i(\theta(M') - \theta(\Gamma)),
\end{align}
where we used Eq.\,\eqref{berryconnection} for $\ell = n-1$ in the second step.
On substituting Eq.\,\eqref{connectionsymmetry} into Eq.\,\eqref{decompose}, 
one obtains
\begin{equation}
\label{integral}
    \int_{M' \to \Gamma \to M} \vec{\mathcal{A}}(\mathbf{k})\cdot {d\mathbf{k}} =
    i\left(\theta(M') - \theta(\Gamma)\right).
\end{equation}
The symmetry of the points $S^{n-1}\Gamma = \Gamma$ and $S^{n-1}M' = M$
enforces the condition that $\ket{u(\Gamma)}$ and $\ket{u(M)}$ are eigenstates of the generator of $U$, namely $\sigma_z$.
We can use this fact to simplify Eq.\,\eqref{integral} to
\begin{equation}
\int_{M' \to \Gamma \to M} \vec{\mathcal{A}}(\mathbf{k})\cdot {d\mathbf{k}} =
    \frac{i\pi m(n-1)}{2n}\left(\braket{\sigma_z}_M - \braket{\sigma_z}_\Gamma\right),
\end{equation}
where $\braket{\sigma_z}_\mathbf{K} = \braket{u(\mathbf{K}) | \sigma_z | u(\mathbf{K})}$ is the spin polarization. 
The same calculation can be repeated for the path $X \to \Gamma \to X'$,
which yields
\begin{equation}
\int_{X \to \Gamma \to X'} \vec{\mathcal{A}}(\mathbf{k})\cdot {d\mathbf{k}} =
    \frac{i\pi m}{2}\left(\braket{\sigma_z}_\Gamma - \braket{\sigma_z}_X\right).
\end{equation}
Adding all contributions and substituting in Eq.\,\eqref{zak-chern} yields 
\begin{equation}
    \frac{\mathcal{C}}{2n} = \frac{m}{4n}\left[\braket{\sigma_z}_\Gamma + (n-1)\braket{\sigma_z}_M - n\braket{\sigma_z}_X\right] \pmod{1}.
\end{equation}
Finally we multiply the whole equation by $2n$ to obtain the 
``Chern-spin polarization relation''
\begin{equation}
    \mathcal{C} = \frac{m}{2}\left[\braket{\sigma_z}_\Gamma + (n-1)\braket{\sigma_z}_M - n\braket{\sigma_z}_X\right] \pmod{2n}.
\label{chern-spin}
\end{equation}
This is the first main result of the present work.

The Chern-spin polarization relation~(\ref{chern-spin}) provides insight into 
the mechanisms responsible for the emergence of energy bands characterized by 
$|\mathcal{C}|=2,3$.
As each $\braket{\sigma_z}_{\mathbf{K}}$ take values in $\{1,-1\}$,
Eq.\,\eqref{chern-spin} implies that $\mathcal{C}$ must be an even integer
for $m=2$ and any value of $n$. Therefore, a Hamiltonian satisfying the $m=2$ 
symmetry after undergoing a topological band inversion typically enters a
$|\mathcal{C}| = 2$ phase. Similarly, for a triangular lattice Hamiltonian 
obeying the $m=3$ symmetry, a topological band inversion typically leads to 
a $|\mathcal{C}| = 3$ phase. Note that a high value of the Chern number, 
i.e.\, $|\mathcal{C}|=2,3$ is not guaranteed, but possible, 
for Hamiltonians obeying the $m=1$ symmetry 
after undergoing topological band inversion. These observations
form a basis for our construction of higher Chern number Hamiltonians 
in the next section.

\subsection{Engineering topological Hamiltonians}
\label{subsec:engineering}

The Chern-spin polarization relation is a powerful tool for engineering 
topological Hamiltonians, as we demonstrate next by constructing a 
tight-binding Hamiltonian with a $|\mathcal{C}|=2$ band. Let $\mathbf{a}_1$ and 
$\mathbf{a}_2$ be the lattice vectors, with angular separation $2\pi/3$, of a 
triangular Bravais lattice, and $\mathbf{a}_3 := -\mathbf{a}_1-\mathbf{a}_2$.
The expression for a general two-band SO coupled Hamiltonian is
\begin{equation}
H(\mathbf{k}) =  h_x \sigma_x + h_y \sigma_y + h_z \sigma_z,
\end{equation}
where $h_x,h_y$, and $h_z$ are $\mathbf{k}$-dependent coefficients, and a
spin-independent term proportional to the identity is omitted as it has no effect 
on the Chern number.
Our strategy is to construct a gapped Hamiltonian which obeys the $m=2$ symmetry
and satisfies $\text{sgn}(h_z(M)) = -\text{sgn}(h_z(\Gamma))$. 
By Eq.\,\eqref{chern-spin}, such a Hamiltonian must have $|\mathcal{C}|=2$.
To restrict the Hamiltonian to nn terms, each of the coefficients 
$h_x,h_y$ and $h_z$ must be
linear combinations of $\{\cos k_q, \sin k_q, \ q=1,2,3\}$, with $k_q = \mathbf{k}\cdot\mathbf{a}_q$.
As $\sigma_z$ satisfies $U\sigma_z U^\dagger = \sigma_z$, the constraint on $h_z$ is $h_z(\mathbf{k}) = h_z(S\mathbf{k})$,
which restricts the possible values to 
\begin{equation}
\label{zcoeff}
        h_z = M_z + 2t_z\textstyle\sum_{q}\cos(k_q), \;\; M_z,t_z \in \mathbb{R}.
\end{equation}
Note that a term proportional to $\textstyle\sum_{q}\sin(k_q)$ in $h_z$ 
breaks the required symmetry of the Hamiltonian. A simple check is 
to note that we require $h_z(S^3\mathbf{k}) = h_z(-\mathbf{k}) = h_z(\mathbf{k})$,
and the latter equality is not obeyed by a term proportional 
to $\textstyle\sum_{q}\sin(k_q)$ in $h_z$. 
The constraints on the coefficients of $\sigma_x$ and $\sigma_y$ imposed by the symmetry lead to
\begin{align}
\label{xycoeff}
    h_x &= 2t_{\rm so}\textstyle\sum_{q}\cos(k_q)\cos(2q\pi/3),\nonumber\\[2pt]
    h_y &= -2t_{\rm so}\textstyle\sum_{q}\cos(k_q)\sin(2q\pi/3), \;\; t_{\rm so} \in \mathbb{R},
\end{align}
which completes the construction of the Hamiltonian. 
We note that for small values of $|\mathbf{k}|$, this exotic SO 
coupling term
\begin{eqnarray}
h_x\sigma_x + h_y\sigma_y&\propto&(k_x^2-k_y^2-2\sqrt{3}k_xk_y)\sigma_x
\nonumber \\
&+&(\sqrt{3}k_x^2-\sqrt{3}k_y^2 + 2k_xk_y)\sigma_y,
\end{eqnarray}
is quadratic in $k_x,k_y$.
After simplification using trigonometric identities, the terms reduce to
\begin{align}
\label{tight-binding}
    h_x &= t_{\rm so}(2\cos(k_3) - \cos(k_1) - \cos(k_2)),\nonumber\\
    h_y &= -\sqrt{3}t_{\rm so}(\cos(k_1) - \cos(k_2)),\nonumber\\
    h_z &= M_z + 2t_z(\cos(k_1) + \cos(k_2) + \cos(k_3)).
\end{align}

\begin{figure}[t]
\includegraphics[width=\columnwidth]{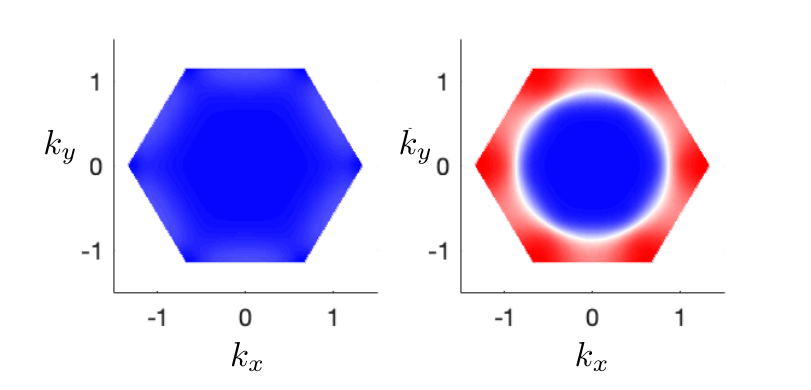}
\caption{(Color online) Color map of spin polarization over the BZ for the 
trivial (left) and the topological (right) phase with~${\mathcal C}=-2$. 
The crystal momenta $k_x$ and $k_y$ are in units of $\pi/a$, with 
$a = |\mathbf{a}_q|$ denoting the
lattice constant. The bright red (blue) color represents $+1$ ($-1$) value of the spin polarization $\braket{\sigma_z}_{\mathbf{k}}$. 
Parameters used were $t_{\rm so} = 0.5, t_z = 1$, 
$M_z=4$ (trivial phase) and $M_z=1$ (topological phase). \label{fig:spindance}}
\end{figure}

For $t_{\rm so}\ne 0$ and $t_z >0$, this Hamiltonian is gapped except for 
$M_z/t_z \in \{-6,2,3\}$. Further restricting to $M_z/t_z > -6$ and using 
$k_q = 0$ and $k_q = -2\pi/3$ at the $\Gamma$ and $M$ points respectively for 
$q \in \{1,2,3\}$, we obtain $\ket{u(\Gamma)} = \ket{\downarrow}$, and 
\begin{equation}
    \ket{u(M)} = \left\{\begin{array}{lcl} \ket{\downarrow} & \text{if} & M_z/t_z > 3,\\
    \ket{\uparrow} & \text{if} & M_z/t_z < 3.
    \end{array}\right.
\end{equation}
Therefore, Eq.\,\eqref{chern-spin} dictates that the lower band 
has Chern number $\mathcal{C} = -2\;(\mbox{mod}\,6)$ 
in the parameter range $-6 < M_z/t_z < 3$ (except $M_z/t_z=2$ at which the gap closes), 
and $0$ otherwise, which we confirm numerically. 
Figure \ref{fig:spindance} shows the variation of the spin polarization
of the lower energy band in the BZ. In the topological phase, the spin 
polarization at the corners of the BZ including at the $M$ point, is the 
negative of the polarization at the center, $\Gamma$. 

One can now construct other Hamiltonians that host $|\mathcal{C}|=2,3$ phases 
by following a similar strategy. A nearest-neighbor two-band Hamiltonian on a 
triangular Bravais lattice with $\mathcal{C}=2$ is obtained by leveraging the 
Chern-spin polarization relation~\eqref{chern-spin} for the $m=1$ symmetry. The 
symmetry condition places the exact same constraints on the $h_z$ coefficient 
as in Eq.~(\ref{zcoeff}), while the constraints on the $\sigma_x$ and 
$\sigma_y$ coefficients lead to
\begin{eqnarray}
h_x &=& 2t_{\rm so}\sum_{q}\sin(k_q)\cos(2q\pi/3);\nonumber \\
h_y &=& 2t_{\rm so}\sum_{q}\sin(k_q)\sin(2q\pi/3).
\label{xycoeff2}
\end{eqnarray}
For small values of $|\mathbf{k}|$, it can be verified that
\begin{equation}
h_x\sigma_x + h_y\sigma_y \propto [S_z(-\pi/6)\mathbf{k}] 
\times \vec{\sigma}\cdot \hat{z},
\end{equation}
which is the Rashba SO coupling. The Chern number can again be evaluated 
directly by using the Chern-spin polarization relation. For $t_{\rm so} \ne 0$ 
and $M_z,t_z>0$, the spin at the $\Gamma$ point is 
$\braket{\sigma_z}_\Gamma = -1$. For $2<M_z/t_z < 3$, the spin polarizations at 
the $M$ and the $X$ points are $\braket{\sigma_z}_M = -\braket{\sigma_z}_X = 1$,
which leads to a $\mathcal{C} = 2$ phase. Note that this Hamiltonian exhibits 
a $|\mathcal{C}|=2$ phase for a smaller parameter range than was the case above
when $m=2$. This could be attributed to the fact that in contrast to the $m=2$ 
symmetric Hamiltonian, the $m=1$ symmetric Rashba SO term does not enforce the 
Chern number to be even in general, and therefore a finer tuning of parameters 
is required.

The construction of tight-binding Hamiltonians with $|\mathcal{C}| = 3$ 
follows a similar approach. Equation (\ref{chern-spin}) for $m=1$ with the 
additional condition $\braket{\sigma_z}_\Gamma = \braket{\sigma_z}_M$ reduces 
to
\begin{equation}
    \mathcal{C} = \frac{n}{2}\big[\braket{\sigma_z}_\Gamma
    -\braket{\sigma_z}_X \big]\;(\mbox{mod}\,2n).
\end{equation}
The additional condition is satisfied for
\begin{align}
\label{zcoeffnnn}
        h_z &= M_z + 2t_z[\cos(k_1-k_2)+\cos(k_2-k_3)\nonumber \\
	&\qquad +\cos(k_3-k_1)].
\end{align}
To satisfy the $m=1$ symmetry, the coefficients of $\sigma_x$ and $\sigma_y$ 
are taken to have the same form as in Eq.~\eqref{xycoeff2}. In the parameter 
range $-6<M_z/t_z<2$ and $t_{\rm so} \ne 0, t_z>0$, one obtains
$\braket{\sigma_z}_\Gamma = -\braket{\sigma_z}_X = -1$, and therefore this 
Hamiltonian displays a $\mathcal{C}=-3$ phase. Next consider the $m=3$ symmetry 
case. Choosing $h_z$ as in Eq.~(\ref{zcoeff}), any two-band nearest-neighbor 
Hamiltonian is necessarily gapless, as $h_x,h_y \propto \sum_q \sin k_q$. One
way to construct $h_x$ and $h_y$ coefficients satisfying the symmetry is to 
add next-nearest-neighbor SO coupling. One such choice is
\begin{eqnarray}
    h_x &=& t_{\rm so}\sum_q \sin k_q;\nonumber \\
    h_y &=& t_{\rm so}[\sin(k_1-k_2)+\sin(k_2-k_3)+\sin(k_3-k_1)].\hphantom{aaa}
\end{eqnarray}
For small values of $|\mathbf{k}|$, this SO coupling has a
complicated cubic dependence on $k_x,k_y$.
For any $t_{\rm so} \ne 0$, the total Hamiltonian displays a $|\mathcal{C}|=3$ 
phase for the parameter range $-6<M_z/t_z<2$. 

We now use our approach for the construction of a $\mathcal{C}=2$ Hamiltonian using NNN terms.
For~$m=2$, we only require $\braket{\sigma_z}_\Gamma = -\braket{\sigma_z}_M$ to ensure $|\mathcal{C}| = 2$. 
This is achieved by isotropic spin-dependent hopping term
\begin{equation}
    h_z = M_z + 2t_z\sum_{q =1,2}\cos k_q, \;\; k_q = \mathbf{k}\cdot \mathbf{a}_q^{\rm sq}
\end{equation}
for $|M_z| < 4|t_z|$, where $\mathbf{a}_1^{\rm sq}, \mathbf{a}_2^{\rm sq}$ are 
the lattice vectors of a square Bravais lattice.
To satisfy the $m=2$ symmetry using nearest neighbor terms, the only possibility is
$h_x ,h_y \propto \cos k_1 - \cos k_2$, which would lead to a gapless Hamiltonian. 
A gapped Hamiltonian can be constructed by introducing a weak diagonal hopping,
\begin{align}
    h_x &= t_{\rm so}(\cos k_1 - \cos k_2), \nonumber\\
    h_y &= t_{\rm so}[\cos (k_1+k_2) - \cos (k_1-k_2)],
\end{align}
which then displays $|\mathcal{C}|=2$ phase. Note the similarity with the $|\mathcal{C}|=3$ Hamiltonian
on a triangular lattice satisfying the $m=3$ symmetry constructed above.

Following our approach, one may also attempt the construction of a square 
lattice Hamiltonian obeying the $m=1$ symmetry with $\mathcal{C}=2$.
Such a Hamiltonian however requires strong next-nearest-neighbor $\sigma_z$ terms, as otherwise
$\braket{\sigma_z}_\Gamma = \braket{\sigma_z}_M$ and $\braket{\sigma_z}_\Gamma = -\braket{\sigma_z}_X$
cannot be satisfied simultaneously. A closely related Hamiltonian with $|\mathcal{C}|=2$ bands 
has been constructed in Ref.\,\cite{sticlet2012}.

The Hamiltonians we construct above enter a higher Chern number topological phase for 
infinitesimal values of the SO coupling strength $t_{\rm so}$, which is 
advantageous for experimental realization. Most of these Hamiltonians host 
$|\mathcal{C}|=2$ or $|\mathcal{C}|=3$ phases for $M_z = 0$, which could be 
important for discovering materials that show quantum anomalous Hall effect in 
the total absence of external magnetic field. The construction of 
other topological Hamiltonians by leveraging symmetries 
for the purpose of ultracold atomic
simulation has also been 
investigated in Ref.\,\cite{kuno2018}.

\section{Realizing and detecting a $|{\mathcal C}|=2$ phase with ultracold atoms}
\label{sec:ultracold}

\subsection{Overview of the proposed experimental setup}
\label{subsec:experiment}

We now describe how a topological Hamiltonian on a triangular lattice, 
described by Eqs.~\eqref{zcoeff} and \eqref{xycoeff}, could be realized using 
ultracold~$^{87}$Rb atoms.  The scheme we present here only requires isolation 
of, and control over, three hyperfine levels forming a $\Lambda$ configuration
for the realization of the SO coupling via Raman coherence, and therefore can 
be extended to all bosonic and fermionic atoms used in ultracold atomic 
experiments. 

\begin{figure}[t]
    \centering
    \includegraphics[width = 0.7\columnwidth]{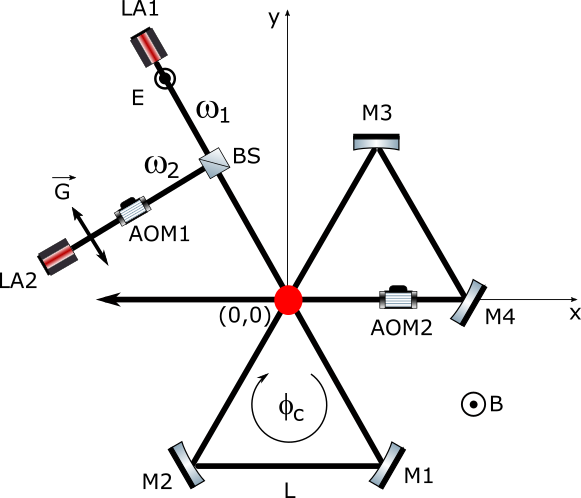}
    \caption{Proposed experimental setup for the realization and the detection of a 
    $|\mathcal{C}| = 2$ topological phase.
    In the figure, LA1 and LA2 denote two lasers with the same frequency $\omega_1$
    polarized out of plane (along $z$ axis) and in plane (in $x$-$y$ plane),
    respectively. BS denotes a $50:50$ beam-splitter, M1--M4 are
    polarization-preserving mirrors, and AOM1 and AOM2 are acousto-optic modulators.
    An external magnetic field of strength $B$ is applied out of plane.}
    \label{fig:scheme}
\end{figure}
By applying a red-detuned laser sheet that is tightly focused 
along the $z$ axis, the atoms are confined to the $xy$ plane 
(Fig.~\ref{fig:scheme}). The atoms are then loaded onto a spin-independent
triangular optical lattice potential~\cite{Becker_2010}, formed by three 
red-detuned $z$-polarized (linear $\pi$-polarized) ``lattice'' laser beams originating
from the laser LA1 in Fig.~\ref{fig:scheme}.
The scheme for inducing SO coupling, inspired by the experimental setup 
described in Ref.~\cite{wu2016}, makes use of two hyperfine levels of the ground-state 
manifold as the pseudospin-$1/2$ up ($\ket{\uparrow}$) and 
down ($\ket{\downarrow}$) states. To simulate SO 
coupling between these two states, a 
sufficiently strong external magnetic field needs to be applied in the 
$z$ direction, so that the hyperfine energy levels shift according to the 
quadratic Zeeman effect. Three ``Raman'' laser beams originating
from the laser LA2 with linear in-plane polarization
induce Raman transitions between the $\ket{\uparrow}$ and $\ket{\downarrow}$ levels 
via excited states. The Raman laser beams also generate an
additional spin-dependent periodic potential.

For the detection of the Chern number, we design a scheme 
based on Bloch oscillations and TOF imaging that leverages
the Chern-spin polarization relation. The spin polarization 
of the Bose-Einstein condensate of bosonic ultracold atoms
can be obtained by releasing the atoms from the trap followed by
a Stern-Gerlach measurement. The condensate can be moved from 
one point in the BZ to another by 
short Bloch oscillations. Such oscillations are induced
by slowly accelerating the optical lattice, which is achieved by slowly 
and simultaneously chirping the frequencies of the lattice and Raman lasers. 
If the experiment is performed with fermionic atoms, 
then the Bloch oscillations are not required, as a direct 
spin-resolved TOF imaging suffices to obtain the spin polarization. 
Finally, the Chern number is 
calculated using the Chern-spin polarization relation.

\subsection{Derivation of the SO coupled Hamiltonian in the continuum}
\label{sec:continuumHam}
\begin{figure}[t]
    \centering
    \includegraphics[width = 0.6\columnwidth]{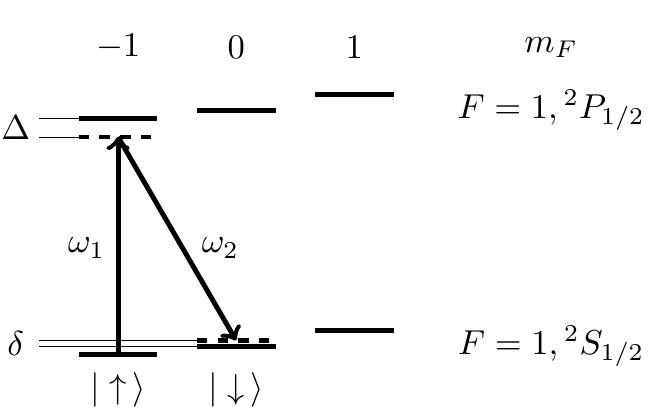}
    \caption{The $\Lambda$-configuration, responsible for the Raman transition, formed 
    by $\ket{\uparrow}$ and $\ket{\downarrow}$ states with
    the excited state $\ket{F=1,m_F=-1}$ in the $2P_{1/2}$ multiplet. $F=2$ levels of the $2P_{1/2}$ multiplet
    are not shown for simplicity.}
    \label{fig:level}
\end{figure}

\subsubsection{Laser configuration}
\label{subsubsec:lasers}

As per convention, the $z$-axis is the quantization axis of the atom, so that 
$J_z\ket{J,m,F,m_F} = \hbar (m + m_F)\ket{J,m,F,m_F}$, where 
$J_z = L_z + S_z + I_z$ is the $z$-component of the total angular momentum 
operator. The pseudospin-$1/2$ up and down states can then be chosen to 
correspond to two hyperfine levels of the ground state manifold, namely 
$\ket{\uparrow}:=\ket{F=1,m_F=-1}$ and $\ket{\downarrow}:=\ket{F=1,m_F=0}$,
as depicted in Fig.~\ref{fig:level}. The three lattice laser beams with 
amplitude~$E_{0}$ propagate in the $xy$-plane with wavevectors
\begin{equation}
    \mathbf{k}_1 = k\left(-1,0\right), \;
    \mathbf{k}_2 = k\left(\frac{1}{2},-\frac{\sqrt{3}}{2}\right), \;
    \mathbf{k}_3 = k\left(\frac{1}{2},\frac{\sqrt{3}}{2}\right).
\end{equation}
The frequency $\omega_1$ of the lattice laser beams is close to, but less than, 
the $D1$ transition frequency. 
The three Raman laser beams are linear polarized in the $xy$ plane and
propagate along the same directions as the lattice laser beams with amplitude 
$G_{0}$, and frequency $\omega_2=\omega_1
+(\epsilon_{\uparrow}-\epsilon_{\downarrow})+\delta$, where 
$\epsilon_\uparrow - \epsilon_{\downarrow}$ is the Zeeman energy shift, and 
$\delta$ denotes the two-photon detuning. The frequency shift can be 
implemented with the help of an acousto-optic modulator AOM1, as shown in 
Fig.~\ref{fig:scheme}. The arm length $L$ needs to be adjusted such that the 
relative phase acquired by adjacent Raman laser beams is 
$\phi_c := 3L\delta\omega/c = 2\pi/3$. 

The combination of the three in-plane 
polarized beams at any point in the $xy$-plane can be separated into 
superpositions of $\sigma_+$ and $\sigma_-$ circular polarization. 
In the spherical basis
\begin{equation}
    \mathbf{e}_{+}=\left(-\frac{1}{\sqrt{2}}, -\frac{i}{\sqrt{2}}\right),
    \quad \mathbf{e}_{-}=\left(\frac{1}{\sqrt{2}}, -\frac{i}{\sqrt{2}}\right),
\end{equation}
and using the definition
    $G_{\pm}(\mathbf{r}) = \mathbf{e}_{\pm}^*\cdot \vec{G}(\mathbf{r})$,
one obtains
\begin{equation}
    {\mathbf{r}}\cdot\vec{G} = -\frac{(x+iy)}{\sqrt{2}}G_+ +  \frac{(x-iy)}{\sqrt{2}}G_-.
\end{equation}
The $\sigma_-$-polarized component at frequency $\omega_2$ in combination with the 
linearly polarized field at frequency $\omega_1$ leads to Raman coherence in 
the $\Lambda$ system formed by $\ket{\uparrow},\ket{\downarrow}$ and 
$\ket{F,-1}$ for each~$F=1,2$ of the $^2{P}_{1/2}$ multiplet 
(Fig.~\ref{fig:level}). All transitions to the level $\ket{F = 1,m_F = 1}$ in 
the $^2{S}_{1/2}$ multiplet are suppressed due to a large two-photon 
detuning as a result of the quadratic Zeeman shift, and therefore can be 
ignored. The detuning $\Delta$ from the center of the D1 transition is large 
compared to both the hyperfine splitting and Zeeman energy shift in the
$^2{P}_{1/2}$ levels.

We now derive the symmetry properties of the electric fields, which will be 
used later to derive the optical lattice potential and the form of SO 
coupling induced by this laser configuration. The hyperfine splitting is much 
smaller than the $D1$ energy gap, so that $\omega_2-\omega_1 \ll \omega_1$;
therefore, $|\mathbf{k}_2| \approx |\mathbf{k}_1| = k$ is a good assumption. 
The $E(\mathbf{r},t)$ is the complex-valued electric field in the $z$ direction, 
and  $\vec{G}(\mathbf{r},t):=(G_x(\mathbf{r},t),
G_y(\mathbf{r},t))$ is the in-plane electric field along the $x$ and $y$ 
directions. One can now write $E(\mathbf{r},t) = E(\mathbf{r})e^{-i\omega_1 t}$,
$\vec{G}(\mathbf{r},t) = \vec{G}(\mathbf{r})e^{-i\omega_2t}$, where
\begin{equation}
\label{fields}
    E(\mathbf{r})  =  E_0 \sum_{j=1}^{3} e^{i\mathbf{k}_j\cdot \mathbf{r}}, \quad
    \vec{G}(\mathbf{r})  =  G_0 \sum_{j=1}^{3} e^{i(\mathbf{k}_j\cdot \mathbf{r} + j\phi_c)}\hat{z}\times \mathbf{k}_j.
\end{equation}
Let \begin{equation}
\label{latticevectors}
    \mathbf{a}_1 = \frac{4\pi}{3k}\left(1,0\right), \; \;\;
    \mathbf{a}_2 = \frac{4\pi}{3k}\left(-\frac{1}{2},\frac{\sqrt{3}}{2}\right).
\end{equation}
The electric fields satisfy
\begin{eqnarray}
E(\mathbf{r} + \mathbf{a}_q) &=& e^{i2\pi/3} E(\mathbf{r});\nonumber \\
G_\pm(\mathbf{r} + \mathbf{a}_q) &=& e^{i2\pi/3} G_\pm(\mathbf{r}),\quad 
q = 1,2,3.
\label{translation}
\end{eqnarray}
Equation \eqref{translation} follows directly from the observation 
$\mathbf{k}_j \cdot \mathbf{a}_q = 2\pi/3 \mod 2\pi$ for all 
$q,j \in \{1,2,3\}$. Therefore, $\mathbf{a}_1, \mathbf{a}_2$ are the lattice
vectors of the triangular optical 
lattice formed by the lattice lasers~\cite{Becker_2010}.
We also define $\mathbf{a}_3 = -\mathbf{a}_1-\mathbf{a}_2$ as before.
The reciprocal lattice vectors are
\begin{equation}
    \mathbf{b}_1 = \sqrt{3}k\,\left(\frac{\sqrt{3}}{2}, \frac{1}{2}\right),
    \quad \mathbf{b}_2 = \sqrt{3}k\,\left(0,1\right).
\end{equation}

For the rotation by $2\pi/3$ around the $z$-axis implemented by
\begin{eqnarray}
R^2 &=& R_z(2\pi/3) = \begin{bmatrix} \cos (2\pi/3) & -\sin (2\pi/3) \\ 
\sin(2\pi/3) & \cos(2\pi/3)\end{bmatrix}\nonumber \\
&=& \begin{bmatrix} -1/2 & -\sqrt{3}/2 \\ \sqrt{3}/2 & -1/2\end{bmatrix},
\end{eqnarray}
    one obtains
     \begin{equation}
     \label{rotation}
        E(R^2\mathbf{r}) = E(\mathbf{r}), \; \;
        G_\pm(R^2\mathbf{r}) = e^{i(\phi_c \mp 2\pi/3)} G_\pm(\mathbf{r}).
        \end{equation}
        Equation \eqref{rotation} follows from the rotational symmetry of the in-plane and out-of-plane field components. The out-of-plane field is invariant under rotation by $2\pi/3$, which leads to the first equality in Eq.\,\eqref{rotation}.
    The second equality can be derived as follows: First note that by symmetry of the
    experimental setup, the field $\vec{G}$ satisfies
    \[
    \vec{G}(R^2\mathbf{r}) = e^{i\phi_c}R^2\vec{G}(\mathbf{r}).
    \]
Using
$\mathbf{e}_{\pm}^* R^2 =  e^{\mp i2\pi/3}\mathbf{e}_{\pm}^*$, one obtains
\begin{align*}
    G_\pm(R^2\mathbf{r}) &= \mathbf{e}_{\pm}^* \cdot \vec{G}(R^2\mathbf{r}) \nonumber \\
     &= e^{i\phi_c}\mathbf{e}_{\pm}^* \cdot R^2\vec{G}(\mathbf{r}) \nonumber \\
      &= e^{i(\phi_c\mp 2\pi/3)}\mathbf{e}_{\pm}^* \vec{G}(\mathbf{r}) \nonumber \\
       &= e^{i(\phi_c \mp 2\pi/3)} G_\pm(\mathbf{r}),
    \end{align*}
    as desired.

\subsubsection{Optical lattice potential}
\label{subsubsec:lattice}

We now compute the optical lattice potential due to the lattice laser beams.
The frequency $\omega_1$ can be chosen to be close to $D1$ transition frequency 
so as to ensure that $D2$ transitions can be safely ignored. This assumption is
made only for simplicity; the scheme works even if both D1 and D2 transitions 
have comparable contributions. The Rabi frequency of oscillation due to the 
optical lattice lasers is
\begin{align}
    \hbar \Omega_{\uparrow, 0,F}(\mathbf{r}) &=\langle F,-1|e z|\uparrow \rangle E(\mathbf{r}), \nonumber\\
    \hbar \Omega_{\downarrow, 0,F}(\mathbf{r}) &=\langle F,0|e z| \downarrow \rangle E(\mathbf{r}),
\end{align}
where $F$ corresponds to the excited level in the $^2{P}_{1/2}$ space.
The optical lattice potential generated by the laser with $\omega_1$ frequency alone
is 
\begin{equation}
    V(\mathbf{r})= \sum_{F=1,2}\frac{\hbar}{4\Delta}
    \left|\Omega_{\sigma,0,F}(\mathbf{r})\right|^{2},
\end{equation}
where $\Delta$ is the detuning. Note that since $\Omega_{\sigma,0,F} \propto 
E(\mathbf{r})$, then $V(\mathbf{r})\propto |E(\mathbf{r})|^2$. The lattice 
potential is generated by $\pi$-polarized light and therefore is 
spin-independent (see discussion above Eq.\,(44) in 
Ref.~\cite{steck2001rubidium} for a rigorous justification).  The details of 
the exact potential can be found in Ref.~\cite{Becker_2010}. The contribution 
to the continuum Hamiltonian due to lattice lasers alone is
\begin{equation}
    H_{\rm lat} = -\frac{\hbar^2\nabla^2}{2m} + V({\mathbf{r}}).
\end{equation}

\subsubsection{Spin-dependent potential}
In contrast to the out-of-plane polarized lattice lasers, the Raman lasers add 
a spin-dependent component to the optical lattice potential.
The dipole potential for alkali atoms in the state $(F,m_F)$ due 
to $D1$ transitions is given by the formula~\cite{GRIMM200095}
\begin{equation}
V_{m_F}(\mathbf{r})=\frac{\pi c^{2} \Gamma}{2 \omega_{0}^{3}}\left(\frac{1-\mathcal{P} g_{F} m_{F}}{\Delta}\right) I(\mathbf{r}),
\end{equation}
where the detuning $\Delta$ is with respect to the center of the $D1$ line,
$g_{F}$ is the Landé factor, $m_{F}$ is the relevant magnetic spin state of 
the atom, and $\mathcal{P}=\pm 1$ and $\mathcal{P}=0$ represent the local 
polarization $\sigma^{\pm}$ and $\pi$, respectively, of the light field 
relative to the chosen quantization axis. For the $\ket{\downarrow}$ state, 
$m_F = 0$, and therefore the second term vanishes. However, for the 
$\ket{\uparrow}$, we have $m_F = -1$ and therefore the second term is non zero.
We define
\begin{equation}
    W(\mathbf{r}) = V_{\uparrow}(\mathbf{r}) - V_{\downarrow}(\mathbf{r}) =
    \frac{\pi c^{2} \Gamma g_{F}}{2 \omega_{0}^{3}\Delta}\sum_{i=\sigma_\pm}\mathcal{P}_i I_i(\mathbf{r})
\end{equation}
as additional spin-dependent dipole potential due to Raman lasers. This contributes the term
$W(\mathbf{r})\sigma_z/2$ to the total Hamiltonian. The 
$V_{\uparrow}(\mathbf{r}) + V_{\downarrow}(\mathbf{r})$ term can be ignored
because it is spin independent and can be absorbed in $V(\mathbf{r})$.

\subsubsection{Spin-orbit coupling}
\label{subsubsec:SO}

The Raman laser beams, together with the lattice laser beams, induce Raman 
coupling between the $\ket{0}$ and $\ket{1}$ states. Both $F=1$ and $F=2$ 
levels in the $^2{P}_{1/2}$ space contribute to Raman resonance. The 
effective two-photon Rabi frequency is
\begin{equation}
    \Omega_{\rm R}(\mathbf{r}) = \frac{1}{2\Delta} \sum_{F=1,2}\Omega_{\downarrow,-,F}^*(\mathbf{r}) \Omega_{\uparrow,0,F}(\mathbf{r}),
\end{equation}
where
\begin{align}
    \hbar \Omega_{\downarrow,-,F}(\mathbf{r}) &= \frac{1}{\sqrt{2}}\langle F,-1 |e (x - iy)| \downarrow\rangle G_-(\mathbf{r}).
\end{align}
Note that the two-photon Rabi frequency satisfies
\[
\Omega_{\rm R}(\mathbf{r}) \propto E(\mathbf{r})G_-^*(\mathbf{r}).
\]
It follows from Eqs.\,\eqref{translation} and \eqref{rotation} that
\begin{align}
\Omega_{\rm R}(\mathbf{r} + \mathbf{a}_q) &= \Omega_{\rm R}(\mathbf{r}), \;\; q=1,2,3,\label{ramantranslation}\\
\Omega_{\rm R}(R^2\mathbf{r}) &= e^{-i(\phi_c + 2\pi/3)}\Omega_{\rm R}(\mathbf{r}). \label{ramanrotation}
\end{align}
Then, the contribution to the Hamiltonian due to Raman resonance is \cite{Brion_2007}
\begin{equation}
    V_{\rm R}(\mathbf{r})=-\hbar\begin{bmatrix}\frac{\delta}{2}+\sum_{F}\frac{\left|\Omega_{\downarrow,0,F}(\mathbf{r})\right|^{2}}{4 \Delta} & \frac{\Omega_{\rm R}^{*}(\mathbf{r})}{2} \\ \frac{\Omega_{\rm R}(\mathbf{r})}{2} &
    -\frac{\delta}{2}+\sum_{F}\frac{\left|\Omega_{\uparrow,-,F}(\mathbf{r})\right|^{2}}{4 \Delta}\end{bmatrix},
\end{equation}
where $\delta$ is the two-photon detuning. Note that the terms $\sum_F\frac{\left|\Omega_{\downarrow,0,F}(\mathbf{r})\right|^{2}}{4 \Delta}$ in the top left and $\sum_F\frac{\left|\Omega_{\uparrow,-}(\mathbf{r})\right|^{2}}{4 \Delta}$ in the bottom right
corner have already been taken into account when calculating the optical lattice potential and the spin-dependent correction to it.

After combining contributions from all factors,
the total effective Hamiltonian in the continuum becomes
\begin{equation}
    H = H_{\rm lat}  + \left(\frac{W(\mathbf{r})}{2} - \frac{\hbar \delta}{2}\right)\sigma_z - \left(\frac{\hbar \Omega_{\rm R}(\mathbf{r})}{2}\sigma^+
    + \text{H.c.}\right),
\end{equation}
where $H_{\rm lat}$ is the sum of the kinetic energy and the lattice potential generated by the lattice lasers,
and the coefficients $W(\mathbf{r})$ and $\Omega_{\rm R}(\mathbf{r})$ depend on
experimentally adjustable parameters, namely the amplitudes $E_{0}, G_{0}$,
and the single-photon and two-photon detunings $\Delta$ and $\delta$ respectively.

\subsection{Derivation of the tight-binding Hamiltonian}
\label{subsec:tight}

The effective tight-binding Hamiltonian is constructed by restricting the continuum 
Hamiltonian to the Wannier states 
of the lowest energy band. We show through a detailed analysis of the symmetry 
properties of these Wannier states that the resulting Bloch Hamiltonian has 
matrix coefficients of the form in Eqs.\,\eqref{zcoeff} and \eqref{xycoeff}.
The key step
involves taking advantage of the symmetry properties of the Wannier functions as
well as the electric field configuration to obtain the phases acquired
by atoms while hopping in various directions accompanied by (possible) spin flips.

Consider the Bloch eigenstates $\psi_{n,\mathbf{k}}$
of the lattice Hamiltonian $H_{\rm lat}$ corresponding to $n$th band and crystal momentum $\mathbf{k}$.
Next, construct the Wannier states
\begin{equation}
 \phi_{n,j_1,j_2}^\sigma(\mathbf{r}) =
 \frac{1}{\sqrt{N}}\sum e^{-i\mathbf{k}\cdot{\mathbf{r}}}\psi_{n,\mathbf{k}}^\sigma(j_1\mathbf{a}_1 + j_2\mathbf{a}_2),
\end{equation}
defined for all pairs of integers $\vec{j}:=(j_1,j_2)$. From here onwards, we will restrict attention to the Wannier functions of the lowest energy band $n=0$,
and assume that the Wannier functions $\phi_{j_1,j_2}^\sigma(\mathbf{r}) := \phi_{0,j_1,j_2}^\sigma(\mathbf{r})$
are symmetric under six-fold rotation generated by $R$, that is
\[
\phi_{\vec{0}}^\sigma(\mathbf{r}) = \phi_{\vec{0}}^\sigma(R\mathbf{r}).
\]
Because the lattice Hamiltonian $H_{\rm lat}$ is spin-independent, one also has
\begin{equation}
    \phi_{j_1,j_2}^\uparrow(\mathbf{r}) = \phi_{j_1,j_2}^\downarrow(\mathbf{r})
\end{equation}
Further, since the lattice potential $V(\mathbf{r})$ is real, it is possible to use a gauge in which
$\psi_{n,-\mathbf{k}}(\mathbf{r}) = \psi_{n,\mathbf{k}}^*(\mathbf{r})$, which leads to the conclusion that
the Wannier functions $\phi^\sigma$ are real-valued. Note that the Wannier functions
have been constructed using the band eigenstates of $H_{\rm lat}$, which is 
a topologically trivial Hamiltonian. Therefore, as long as a continuous gauge
for the eigenstates $\psi_{n,\mathbf{k}}$ is chosen, the Wannier functions
are expected to be exponentially localized~\cite{monaco2018}.

The lattice Hamiltonian term $H_{\rm lat}$ leads to isotropic hopping without 
spin-flip with strength $t_{\rm hop}$,
\begin{equation}
     \widehat{H}_{\rm hop} = -t_{\rm hop}\sum_{\braket{\vec{j},\vec{j}'}} 
     \hat{\Phi}_{\vec{j}}^\dagger \hat{\Phi}_{\vec{j}'},
\label{eq:Hhop}
\end{equation}
where $\braket{\vec{j},\vec{j}'}$ denote the restriction that $\vec{j},\vec{j}'$ are
nearest-neighbor sites on the triangular lattice, $\hat{\Phi}_{\vec{j}}^\dagger =
\begin{bmatrix}\hat{\phi}_{\vec{j},\uparrow}^\dagger & \hat{\phi}_{\vec{j},\downarrow}^\dagger \end{bmatrix}$ are arrays of creation
operators corresponding to the Wannier orbitals, and
\begin{equation}
\label{hopping}
    -t_{\rm hop} = \int d^2{\mathbf{r}} (\phi_{\vec{0}}^\downarrow)^*(\mathbf{r}) H_{\rm lat}(\mathbf{r})
    \phi_{\vec{a}_1}^\uparrow(\mathbf{r}),
\end{equation}
where $\vec{a}_1 := (1,0)$, $\vec{a}_2 := (0,1)$, and $\vec{a}_3 := (-1,-1) = -\vec{a}_1-\vec{a}_2$
are the integer coordinates of lattice vectors $\mathbf{a}_1, \mathbf{a}_2$, and $\mathbf{a}_3$ in the units of
$\mathbf{a}_1$ and $\mathbf{a}_2$, and $\vec{0} := (0,0)$.
Note that the Wannier functions are real-valued, and so Eq.\,\eqref{hopping} would still hold if
$(\phi_{0,0}^\downarrow)^*$ was replaced by $\phi_{0,0}^\downarrow$ in the integrand.

The optical lattice formed by a $z$-polarized laser has a spin-independent 
correction due to the in-plane polarized lasers. We claim without proof that 
the correction to the hopping strength due to this contribution is also 
isotropic. Whether this claim is true or false does not affect further 
analysis, because the Wannier basis of the optical lattice potential is 
governed by $z$-polarized lasers, and the spin-independent correction only adds 
a factor proportional to the identity to the Hamiltonian in momentum space
that does not affect the eigenvectors or their topology.

Next consider the contribution to the tight-binding Hamiltonian due to the term 
$W(\mathbf{r})\sigma_z/2$ in the continuum Hamiltonian. Once again, the term 
leads to an equal $\sigma_z$-type hopping between all nearest neighbors,
\begin{equation}
     \widehat{H}_{\rm zhop,1} = t_z\sum_{\braket{\vec{j},\vec{j}'}} \hat{\Phi}_{\vec{j}}^\dagger \sigma_z \hat{\Phi}_{\vec{j}'},
\label{eq:Hzhop1}
\end{equation}
where
\begin{equation}
    t_z := \frac{1}{2}\int d^2{\mathbf{r}} (\phi_{\vec{0}}^\downarrow)^*(\mathbf{r})W(\mathbf{r})
    \phi_{\vec{a}_1}^\uparrow(\mathbf{r})
\end{equation}
Apart from a $\sigma_z$-hopping term, $W(\mathbf{r})\sigma_z/2$ also contributes an on-site
$\sigma_z$ mass term, given by
\begin{equation}
     \widehat{H}_{\rm zhop,2} = m_z\sum_{\vec{j}} \hat{\Phi}_{\vec{j}}^\dagger \sigma_z \hat{\Phi}_{\vec{j}},
\label{eq:Hzhop2}
\end{equation}
where
\begin{equation}
    m_z := \frac{1}{2}\int d^2{\mathbf{r}} (\phi_{\vec{0}}^\downarrow)^*(\mathbf{r})W(\mathbf{r})
    \phi_{\vec{0}}^\uparrow(\mathbf{r})
\end{equation}

The Raman process is responsible for inducing hopping with spin-flip in the tight-binding Hamiltonian.
The amplitude of the hopping term between the Wannier centers
$(0,0)$ and $(j_1,j_2)$ can be computed using the formula
\begin{equation}
    t_{\rm so}(j_1,j_2) = \frac{\hbar}{2}\int d^2{\mathbf{r}}
    (\phi_{\vec{0}}^\downarrow)^*(\mathbf{r})\Omega_{\rm R}(\mathbf{r})
    \phi_{j_1,j_2}^\uparrow(\mathbf{r}).
\end{equation}
Because $\phi^\uparrow = \phi^\downarrow$ and $\phi^\sigma$ is real-valued, one
obtains
\begin{align}
    t_{\rm so}(-\vec{a}_q) & =
    \frac{\hbar}{2}\int d^2{\mathbf{r}} (\phi_{\vec{0}}^\downarrow)^*(\mathbf{r})\Omega_{\rm R}(\mathbf{r})
    \phi_{-\vec{a}_q}^\uparrow(\mathbf{r}) \nonumber\\
    & = \frac{\hbar}{2}\int d^2{\mathbf{r}} (\phi_{\vec{0}}^\downarrow)(\mathbf{r})\Omega_{\rm R}(\mathbf{r})
    (\phi_{-\vec{a}_q}^\uparrow)^*(\mathbf{r}) \nonumber\\
    & = \frac{\hbar}{2}\int d^2{\mathbf{r}} (\phi_{-\vec{a}_q}^\uparrow)^*(\mathbf{r})\Omega_{\rm R}(\mathbf{r})
    \phi_{\vec{0}}^\uparrow(\mathbf{r}) \nonumber\\
    & = \frac{\hbar}{2}\int d^2{\mathbf{r}} (\phi_{-\vec{a}_q}^\downarrow)^*(\mathbf{r})\Omega_{\rm R}(\mathbf{r})
    \phi_{\vec{0}}^\uparrow(\mathbf{r}) \nonumber\\
    & = \frac{\hbar}{2}\int d^2{\mathbf{r}} (\phi_{\vec{0}}^\downarrow)^*(\mathbf{r})\Omega_{\rm R}(\mathbf{r})
    \phi_{\vec{a}_q}^\uparrow(\mathbf{r}) \nonumber\\
    & = t_{\rm so}(\vec{a}_q).
\end{align}
Using $\mathbf{a}_3 = R^2\mathbf{a}_2 = R^4\mathbf{a}_1$ along with the 
translation invariance of the Wannier functions $\phi_{j_1,j_2}(\mathbf{r}) 
= \phi_{\vec{0}}(\mathbf{r}-j_1\mathbf{a}_1-j_2\mathbf{a}_2)$, one obtains
\begin{eqnarray}
    \phi_{\vec{a}_2}^\sigma(R^2\mathbf{r}) 
    &=& \phi_{\vec{0}}^\sigma(R^2\mathbf{r}- \mathbf{a}_2) 
    = \phi_{\vec{0}}^\sigma(R^2(\mathbf{r}- \mathbf{a}_1))
    = \phi_{\vec{0}}^\sigma(\mathbf{r}- \mathbf{a}_1)\nonumber \\
    &=& \phi_{\vec{a}_1}^\sigma(\mathbf{r})
\label{wannierrotation}
\end{eqnarray}
and
\begin{eqnarray}
    \phi_{\vec{a}_3}^\sigma(R^2\mathbf{r})
    &=& \phi_{\vec{0}}^\sigma(R^2\mathbf{r}- \mathbf{a}_3) 
    = \phi_{\vec{0}}^\sigma(R^2(\mathbf{r}- \mathbf{a}_2))
    = \phi_{\vec{0}}^\sigma(\mathbf{r}- \mathbf{a}_2)\nonumber \\
    &=& \phi_{\vec{a}_2}^\sigma(\mathbf{r}).
\end{eqnarray}
The hopping amplitudes $t_{\rm so}(\vec{a}_2)$ and $t_{\rm so}(\vec{a}_3)$ can now be calculated as follows:
\begin{align}
    t_{\rm so}(\vec{a}_2) & = \frac{\hbar}{2}\int d^2{\mathbf{r}} (\phi_{\vec{0}}^\downarrow)^*(\mathbf{r})\Omega_{\rm R}(\mathbf{r})
    \phi_{\vec{a}_2}^\uparrow(\mathbf{r}) \\
    & = \frac{\hbar}{2}\int d^2{\mathbf{s}} (\phi_{\vec{0}}^\downarrow)^*(R^2\mathbf{s})\Omega_{\rm R}(R^2\mathbf{s})
    \phi_{\vec{a}_2}^\uparrow(R^2\mathbf{s}) \\
    & = \frac{\hbar}{2}\int d^2{\mathbf{s}} (\phi_{\vec{0}}^\downarrow)^*(\mathbf{s}) e^{-i(\phi_c+2\pi/3)}\Omega_{\rm R}(\mathbf{s})
    \phi_{\vec{a}_1}^\uparrow(\mathbf{s}) \\
    & = e^{-i(\phi_c+2\pi/3)}t_{\rm so}(\vec{a}_1),
\end{align}
where the change of variable $\mathbf{r} = R^2\mathbf{s}$ is employed in the first step and Eqs.\,\eqref{ramanrotation} and
\eqref{wannierrotation} are used in the later steps. Note that since $R$ is a rotation, the Jacobian of the transformation
is $1$. A similar calculation yields
\[
t_{\rm so}(\vec{a}_3) = e^{-i(\phi_c+2\pi/3)}t_{\rm so}(\vec{a}_2)
\]

The on-site spin-flipping term does not survive for $\phi_c = 2\pi/3$. This follows from the relation
\begin{align}
    t_{\rm so}(\vec{0}) & = \frac{\hbar}{2}\int d^2{\mathbf{r}} (\phi_{\vec{0}}^\downarrow)^*(\mathbf{r})\Omega_{\rm R}(\mathbf{r})
    \phi_{\vec{0}}^\uparrow(\mathbf{r}) \\
    & = \frac{\hbar}{2}\int d^2{\mathbf{s}} (\phi_{\vec{0}}^\downarrow)^*(R^2\mathbf{s})\Omega_{\rm R}(R^2\mathbf{s})
    \phi_{\vec{0}}^\uparrow(R^2\mathbf{s}) \\
    & = \frac{\hbar}{2}\int d^2{\mathbf{s}} (\phi_{\vec{0}}^\downarrow)^*(\mathbf{s}) e^{-i(\phi_c+2\pi/3)}\Omega_{\rm R}(\mathbf{s})
    \phi_{\vec{0}}^\uparrow(\mathbf{s}) \\
    & = e^{-i(\phi_c+2\pi/3)}t_{\rm so}(\vec{0}),
\end{align}
which implies that $t_{\rm so}(\vec{0}) = 0$ for $\phi_c =2\pi/3$.

One can redefine $\phi^\downarrow$ by a global gauge transformation such that 
$t_{\rm so}(\vec{a}_3)$ is real and positive. From here onwards, we set 
$t_{\rm so} := t_{\rm so}(\vec{a}_3)$.
Now the total contribution to the tight-binding Hamiltonian due to the Raman process is
\begin{eqnarray}
    \widehat{H}_{\rm so} &=&
    \sum_{\vec{j}}\sum_{q=1}^{3}t_{\rm so}\Big\{\hat{\Phi}_{\vec{j}}^\dagger 
    \big[\cos (-2q\pi/3)  \sigma_x\nonumber \\
    &&\qquad\qquad + \sin (-2q\pi/3) \sigma_y\big] \hat{\Phi}_{\vec{j} +
    \vec{a}_q} + \text{H.c.}\Big\}.\hphantom{aaa}
\label{eq:Hso}
\end{eqnarray}
Finally, the two-photon detuning leads to on-site $\sigma_z$ term with strength $-\hbar \delta \sigma_z/2$ independent of
the value of the phase difference $\phi_c$,
\begin{equation}
         \widehat{H}_{\rm det} = -\frac{\hbar\delta}{2}\sum_{\vec{j}} \hat{\Phi}_{\vec{j}}^\dagger \sigma_z \hat{\Phi}_{\vec{j}},
\label{eq:Hdet}
\end{equation}
It is safe to assume that the two-photon detuning can be adjusted to be
small enough so that any resulting nn hopping can be ignored.

Combining all terms, one obtains the full tight-binding Hamiltonian
Eqs.~(\ref{eq:Hhop}), (\ref{eq:Hzhop1}), (\ref{eq:Hzhop2}), (\ref{eq:Hso}),
and (\ref{eq:Hdet}). Because the Hamiltonian is number conserving, one may 
write the single-particle Hamiltonian as
\begin{equation}
H = H_{\rm hop} + H_{\rm zhop} + H_{\rm so} + H_{\rm det},
\end{equation}
where
\begin{align}
    H_{\rm hop} &= -t\sum_{\vec{j},q} \left(\ket{\vec{j}} \bra{\vec{j} + \vec{a}_q} + \text{H.c.}\right),
    \nonumber\\
    H_{\rm zhop} &= t_z\sum_{\vec{j},q} \left(\ket{\vec{j}}\bra{\vec{j} + \vec{a}_q} \sigma_z  + \text{H.c.}\right),
    \nonumber\\
    H_{\rm so} &=
    t_{\rm so}\sum_{\vec{j},q}\Big\{\ket{\vec{j}}\bra{\vec{j} + \vec{a}_q}\big[
    \cos (-2q\pi/3)  \sigma_x\nonumber \\
    &\qquad\qquad + \sin (-2q\pi/3)\sigma_y\big]
     + \text{H.c.}\Big\},
     \nonumber\\
    H_{\rm det} &= \left(m_z-\frac{\hbar\delta}{2}\right)\sum_{\vec{j}}\ket{\vec{j}}\bra{\vec{j}} \sigma_z
\end{align}
To obtain the tight-binding Hamiltonian in momentum space, define
\begin{equation}
    V_q = \sum_{\vec{j}}\ket{\vec{j}}\bra{\vec{j}+\vec{a}_q},\quad q=1,2,3,
\end{equation}
and the momentum states
\begin{equation}
    \ket{\mathbf{k}} = \frac{1}{\sqrt{N}}\sum_{\vec{j}} e^{i\mathbf{k}\cdot\mathbf{j}}\ket{\vec{j}}
\end{equation}
where again $\vec{j} = (j_1,j_2)$ and $\mathbf{j} 
= j_1\mathbf{a}_1 + j_2\mathbf{a}_2$. It is easy to verify that
\begin{equation} \label{shift}
   V_q\ket{\mathbf{k}} = e^{ik_q}\ket{\mathbf{k}}, \quad V_q^\dagger\ket{\mathbf{k}} = e^{-ik_q}\ket{\mathbf{k}}.
\end{equation}
After omitting the term proportional to identity in spin space, and
$M_z := m_z-\frac{\hbar\delta}{2}$, one obtains the SO Hamiltonian 
in momentum space 
$H(\mathbf{k}) = \mathbf{H}(\mathbf{k})\cdot\boldsymbol{\sigma}
=h_x \sigma_x + h_y \sigma_y + h_z \sigma_z$, with
the coefficients $h_x,h_y,h_z$ as in Eqs.\,\eqref{zcoeff} and \eqref{xycoeff}.
This is the second main result of the present work.

We conclude the description of the experimental scheme for the 
realization of SO coupling and the desired topological state on a 
triangular lattice by a qualitative comparison
to the scheme used in Ref.\,\cite{wu2016}.  
Despite the apparent similarity, our scheme differs substantially 
from the scheme used in Ref.\,\cite{wu2016} for the square lattice. The lattice 
potential in our scheme is generated by $\pi$-polarized lasers, which are truly 
spin-independent. More important, the Raman potential in our scheme has the 
same periodicity as the optical lattice in the standard gauge. In contrast, the 
scheme in Ref.~\cite{wu2016} leads to a Raman potential that has twice the 
periodicity of the optical lattice before a particular gauge transformation is 
implemented. Such a gauge transformation is not possible in a triangular 
lattice due to the lack of sublattice symmetry. We turn to  
the detection of Chern number in the next section.

\subsection{Chern number via Zeeman Spectroscopy and Bloch oscillations}
\label{subsec:chern}

We now show how the Chern-spin polarization relation, \eqref{chern-spin}, can 
be leveraged to obtain the Chern number (mod 6) of the lower band wavefunction.
The triangular SO-coupled lattice requires $n=3$ and $m=2$, and the relation 
can be further simplified to 
$\mathcal{C}=2(\braket{\sigma_z}_M-\braket{\sigma_z}_\Gamma)~(\text{mod}\,6)$.
Typically, the spin-independent hopping term 
$-2t_{\rm hop}[\cos(k_1) + \cos(k_2) + \cos(k_3)]$ of the Hamiltonian, which 
was omitted in Eq.~\eqref{tight-binding}, dominates as far as the energy 
eigenvalues are concerned. Therefore the~$^{87}$Rb atoms condense to form a 
Bose-Einstein condensate (BEC) at the $\Gamma$ point in the center of the BZ 
that minimizes this term, as shown in Fig.~\ref{fig:loops}. The ratio of the 
populations in the $\ket{\uparrow}, \ket{\downarrow}$ levels obtained by 
Stern-Gerlach imaging (i.e.\ Zeeman spectroscopy) can be used to infer the spin 
polarization at the $\Gamma$ point.

The spin polarization at the point $M$ in the BZ can be obtained by first 
performing a short Bloch oscillation to move the BEC 
adiabatically~\cite{dahan96} from the $\Gamma$ point to the $M$ point. 
For a triangular lattice, the coordinates of the $M$ point, shown in
Fig.~\ref{fig:loops}, are given by
\begin{eqnarray}
    M &=& (-\mathbf{b}_1 + 2\mathbf{b}_2)/3  = \sqrt{3}k\,\left[-\frac{1}{3}\left(0,1\right) +
    \frac{2}{3}\left(\frac{\sqrt{3}}{2}, \frac{1}{2}\right) \right]\nonumber \\
&=& (k,0).
\end{eqnarray}
To map the condensate from the $\Gamma$ point to the $M$ point, it suffices to 
accelerate the lattice along the $x$ direction with some magnitude $a$, which 
is accomplished by varying the frequency of the laser beams 
travelling in the $-x$ direction. Recall that when all three $\omega_1$ beams 
meet at the origin $(0,0)$ in phase, one of the lattice sites coincides with 
the origin, taken to be the center of the lattice. How does the center of the 
lattice shift when the third beam (propagating along $-x$) reaches the origin 
with a phase difference $\varphi$ added by AOM2 with respect to the initial 
configuration? Due to the symmetry of the system, the center must shift to
a point $\mathbf{r}_c = (r,0)$ along the $x$-axis, satisfying
\begin{equation}
    \varphi + \mathbf{k}_1 \cdot \mathbf{r}_c = \mathbf{k}_2 \cdot \mathbf{r}_c = \mathbf{k}_3 \cdot \mathbf{r}_c.
\end{equation}
The second equality is automatically satisfied for any point $\mathbf{r}_c$ along
the $x$-axis. Solving the first equality leads to
\begin{equation}
    \varphi + k\left(-1,0\right)\cdot(r,0) =  k\left(\frac{1}{2},-\frac{\sqrt{3}}{2}\right)\cdot(r,0),
\end{equation}
which gives $r = 2\varphi/3k$.
To achieve acceleration $a$ along the $x$-direction, one needs
\begin{equation}
    \varphi(t) = \frac{3k}{2}\left(\frac{1}{2}a t^2\right) = \frac{3ka t^2}{4}.
\end{equation}
Assuming that the frequency variation is applied at $t=0$, the condition can be 
translated to
\begin{equation}
    \int_{t=0}^{t}\Delta\omega(t)dt = \varphi(t),
\end{equation}
which on differentiating yields the frequency difference as a function of the 
time required to achieve the phase difference $\varphi(t)$ at the origin,
\begin{equation}
    \Delta\omega(t) = \left(\frac{3ka}{2}\right)t.
\end{equation}
Therefore, to achieve acceleration $\alpha$ along $+x$ direction, the frequency 
of the third beam needs to be changed at a linear rate, with 
$d\Delta\omega/dt = 3ka/2$.

The time required to move the BEC from the $\Gamma$ point to the $M$ point can 
be calculated as follows. The rate of change of crystal momentum is given by
\begin{equation}
 \hbar\frac{d\mathbf{k}}{dt} = -ma,
\end{equation}
so that to reach $(k,0)$ from (0,0), the time required is
\begin{equation}
    T = \frac{\hbar k}{ma},
\end{equation}
where $m$ is the mass of the atom.
The acceleration $a$ can be chosen to be arbitrary, but the adiabaticity 
condition~\cite{messiah1962quantum} must be satisfied,
\begin{equation}
    T \gg \frac{\hbar \|dH/dt\|}{\Delta E^2}.
\end{equation}
Typically, we expect the SO-coupling strength $t_{\rm so} \ll t_{\rm hop}$,
so that the energy gap is determined by $t_{\rm so}$ and the bandwidth
by $t_{\rm hop}$. The condition on $a$ then becomes
\begin{equation}
    \frac{\hbar k}{ma} \gg \frac{\hbar t_{\rm hop}}{t_{\rm so}^2} \implies a \ll  \frac{kt_{\rm so}^2}{m t_{\rm hop}}.
\end{equation}

If the Hamiltonian in Eq.~\eqref{tight-binding} is realized using fermionic atoms, and assuming that the density of atoms is adjusted
so that the band is half-filled, then the Chern-spin polarization relation can 
still be 
leveraged to measure the Chern number experimentally. In contrast to bosons, the fermions at half-filling occupy the entire lower energy band due to Fermi-Dirac statistics. 
The spin polarization at both the $\Gamma$ and the $M$ points
can then be directly obtained by standard TOF Stern-Gerlach imaging, which 
involves first turning off the lattice and Raman lasers to let the atoms 
evolve freely for a time $t$ in an external magnetic field, and then imaging 
separately the population of each hyperfine level. At the end of the time 
interval, the atoms with crystal momentum $\mathbf{k}$ reach approximately 
the point $\mathbf{k}t/m\hbar$ in real space, so that the spin amplitude 
at $\mathbf{k}$ can be obtained from the population
difference at the point $\mathbf{k}t/m\hbar$ in real space. 
In this case, the spatial profile of the population difference between the two levels will 
resemble the pattern in Fig.\,\ref{fig:spindance}.

\section{Conclusions}
\label{sec:conclusions}

In this work, we have shown that the origin of some $|\mathcal{C}| = 2,3$ 
phases can be traced to combined real space-spin rotation symmetries for 
SO-coupled lattice gases. This insight leads to a Chern-spin polarization 
relation~(\ref{chern-spin}) that allows for the determination of the Chern 
number by measuring the particle spin polarization at only a small number of
points in the BZ. In the simple setting of nearest-neighbor 
hopping, we demonstrated that triangular Bravais lattices can accommodate 
higher Chern numbers compared to their square counterparts.  We leveraged this
result to provide a detailed proposal for the experimental realization of a 
$|{\mathcal C}|=2$ phase on a triangular lattice using Raman-induced SO 
coupling in ultracold atomic gases. The trivial and 
topological phases can then be distinguished using TOF Zeeman imaging for 
fermions and a combination of Bloch oscillations and TOF Zeeman imaging for 
bosons. Our scheme for the detection of the Chern number suggests that Bloch 
oscillations and Zeeman spectroscopy could be adapted to a large class of SO-coupled 
systems for the detection of Chern insulators in ultracold atoms.

The Chern-spin polarization relation and the tight-binding models that we 
constructed illustrate that symmetries with larger values of $m$ induce 
favorable conditions for the realization of higher Chern number states, which 
should galvanize the search for such states in systems with unconventional SO 
coupling. In the present analysis, the values of $|\mathcal{C}|$ are restricted 
to~$2,3$ due to the smaller cardinality of the symmetry groups that divide the 
BZ in $4$ or $6$ equal parts (c.f.\ Fig.~\ref{fig:loops}). 

To achieve yet higher values of the Chern number, it would be natural to 
explore systems with 
larger symmetry groups, such as those supported on non-Bravais lattices (i.e.\,
Bravais lattices with attached basis)~\cite{holler2018}, systems with 
anti-unitary symmetries~\cite{fu2007,liu2013}, and beyond. We hope to pursue 
these investigations in future work.
It remains to be seen how our methods can be extended to multiband systems, 
including time-reversal invariant systems which have vanishing Chern number 
and systems with intrinsic topological order. Although Zak-Chern relations may 
be trivial for such systems, the topology can often be inferred from the 
wave function at highly symmetric crystal momenta~\cite{fu2007}.

\begin{acknowledgements}
This research was supported by  
the Natural Sciences and Engineering Research Council of Canada and the Alberta Major Innovation Fund.
A.\ A.\ acknowledges support through a Killam 2020 Postdoctoral Fellowship.
\end{acknowledgements}

\bibliographystyle{apsrev.bst}
\bibliography{references}{}

\end{document}